\documentclass[runningheads]{llncs}
\usepackage[T1]{fontenc}
\usepackage{amsmath,amssymb,mathtools,booktabs,longtable,array,tabularx,graphicx}
\makeatletter

\let\c@theorem\relax

\makeatother
\usepackage{amsthm}
\newtheorem{theorem}{Theorem}
\newtheorem{lemma}[theorem]{Lemma}
\newtheorem{corollary}[theorem]{Corollary}
\newtheorem{proposition}[theorem]{Proposition}
\theoremstyle{definition}
\newtheorem{definition}[theorem]{Definition}

\usepackage{tikz,float}
\usetikzlibrary{positioning,arrows.meta}
\usepackage[hidelinks]{hyperref}
\newcommand{\poly}{\operatorname{poly}}

\newcommand{\F}{\mathbb F_2}

\title{Circuit Lower Bounds for Extended Gate Bases from Subexponential Counting}
\titlerunning{Circuit Lower Bounds for Extended Gate Bases}
\author{Mengyang Wu \and Minghao Wang}
\authorrunning{Mengyang Wu and Minghao Wang}
\institute{}
\begin{document}
\raggedbottom
\maketitle
\begin{abstract}
We study circuit lower bounds for extended gate bases using exact CNF counting and an extension-ring form of the contraction theorem of OpenAI.
The deterministic counting theorem of Claude (Anthropic) yields an
$\mathsf E^{\mathsf{NP}}$ language requiring
$\Omega(m\sqrt{\log m}/\log\log m)$ gates over unbounded AND/OR with
binary XOR, infinitely often.
Using that contraction theorem, a bounded binary base with a linear
budget of arbitrary XOR and symmetric gates requires
$\Omega(m\sqrt{\log m})$ gates infinitely often.
These results extend the binary-gate lower bound of OpenAI.
Signed long-clause selectors reduce the first counting problem to the
difference of two ordinary 3-CNF counts.
Affine characters and unit-separated interpolation give the second
counting interface, with logarithmic cost per symmetric-gate arity.
A quantitative local-reduction transfer accounts for auxiliary variables,
branch counts, and integer recovery.
\end{abstract}
\section{Introduction}
\subsection{Problem and motivation}
Faster circuit counting can yield lower bounds for explicit languages through the algorithm-to-lower-bound paradigm. Extending the gate basis changes the cost of verifying a guessed witness circuit. Long defining clauses and high-arity symmetric relations must be represented without exhausting the quantitative saving of the counter.
\subsection{Main results}
Let $\mathcal U$ consist of arbitrary-fan-in AND/OR, binary XOR, and unary NOT/COPY gates, with signed edges and constants allowed. Size counts noninput gates. The first result uses the exact CNF counting theorem of Claude (Anthropic), stated as Theorem~\ref{thm:counting}.
\begin{theorem}\label{thm:unbounded-lower}
There are a language $W\in\mathsf E^{\mathsf{NP}}$
and $c>0$ such that, at infinitely many input lengths $m$, every
$\mathcal U$ circuit computing $W$ has more than
\[
 c\,m\frac{\sqrt{\log m}}{\log\log m}
\]
noninput gates. No depth restriction is imposed.
\end{theorem}
For the second model, ordinary binary gates are supplemented by arbitrary-fan-in XOR and other symmetric gates. Define $b(C)$ by charging one per designated XOR gate and $\log_2(d_j+1)$ per other designated symmetric gate of arity $d_j$. Binary gates count toward size and have zero exceptional budget. Gate lists contain no duplicate signed wires; depth and fanout are unrestricted.
\begin{theorem}\label{thm:symmetric-lower}
Using the contraction theorem of OpenAI as extended in Theorem~\ref{thm:contraction}, there are constants $\delta,c>0$ and
a language $W\in\mathsf E^{\mathsf{NP}}$ such that, for infinitely
many input lengths $m$, every circuit for $W$ with $b(C)\le\delta m$
has at least $c\,m\sqrt{\log m}$ gates. Depth and fanout are unrestricted.
Consequently, for a sufficiently small fixed $\epsilon>0$, the same
lower bound allows simultaneously $\lfloor\epsilon m\rfloor$
arbitrary-fan-in XOR gates and
$\lfloor\epsilon m/\log_2(m+2)\rfloor$ other arbitrary symmetric
gates, with all remaining gates bounded binary.
\end{theorem}
\begin{table}[tbp]\centering\small
\begin{tabularx}{\linewidth}{@{}p{.24\linewidth}X@{}}\toprule
Source / task & Bound and scope \\
\midrule
OpenAI~\cite[Corollary 1.4 and Appendix C.1]{source} & Binary gates: $\Omega(m\sqrt{\log m})$ infinitely often in $\mathsf E^{\mathsf{NP}}$. \\
Theorem~\ref{thm:unbounded-lower} & Unbounded AND/OR and binary XOR: $\Omega(m\sqrt{\log m}/\log\log m)$ infinitely often. \\
Theorem~\ref{thm:symmetric-lower} & Binary base with $b(C)\le\delta m$: $\Omega(m\sqrt{\log m})$ infinitely often, using the extension-ring form of D.1. \\
\bottomrule\end{tabularx}
\caption{Gate models and quantitative lower bounds. The gain is the extended basis supported by each counting interface.}
\end{table}
\subsection{Technical overview}
The proof follows the algorithm-to-lower-bound paradigm: a sufficiently fast exact counter can verify a succinctly represented satisfying assignment faster than a nondeterministic time hierarchy permits. Extending the circuit basis changes the verifier's gate relations. The central task is to count the resulting assignments while charging every auxiliary variable, algebraic branch, and recovery operation. The transfer theorem separates this task from the hierarchy argument, and the two counting interfaces instantiate it with different costs.

\begin{figure}[H]\centering
\begin{tikzpicture}[node distance=5mm,box/.style={draw,rounded corners,align=center,text width=.83\linewidth,inner sep=5pt,font=\small}]
\node[box] (a) {Local decoder and guessed witness circuit};
\node[box,below=of a] (b) {I: signed clauses $\to$ two 3-CNF counts};
\node[box,below=of b] (c) {II: affine characters + unit interpolation $\to$ weighted contractions};
\node[box,below=of c] (d) {Full counting-time bounds $\to$ quantitative transfer};
\node[box,below=of d] (e) {Hierarchy contradiction $\to$ the two circuit lower bounds};
\draw[->] (a)--(b);\draw[->] (a.east) -- ++(.25,0) |- (c.east);\draw[->] (b.west)-- ++(-.25,0) |- (d.west);\draw[->] (c)--(d);\draw[->] (d)--(e);
\end{tikzpicture}
\caption{Two counting interfaces feed the same quantitative transfer theorem.}\label{fig:overview}
\end{figure}
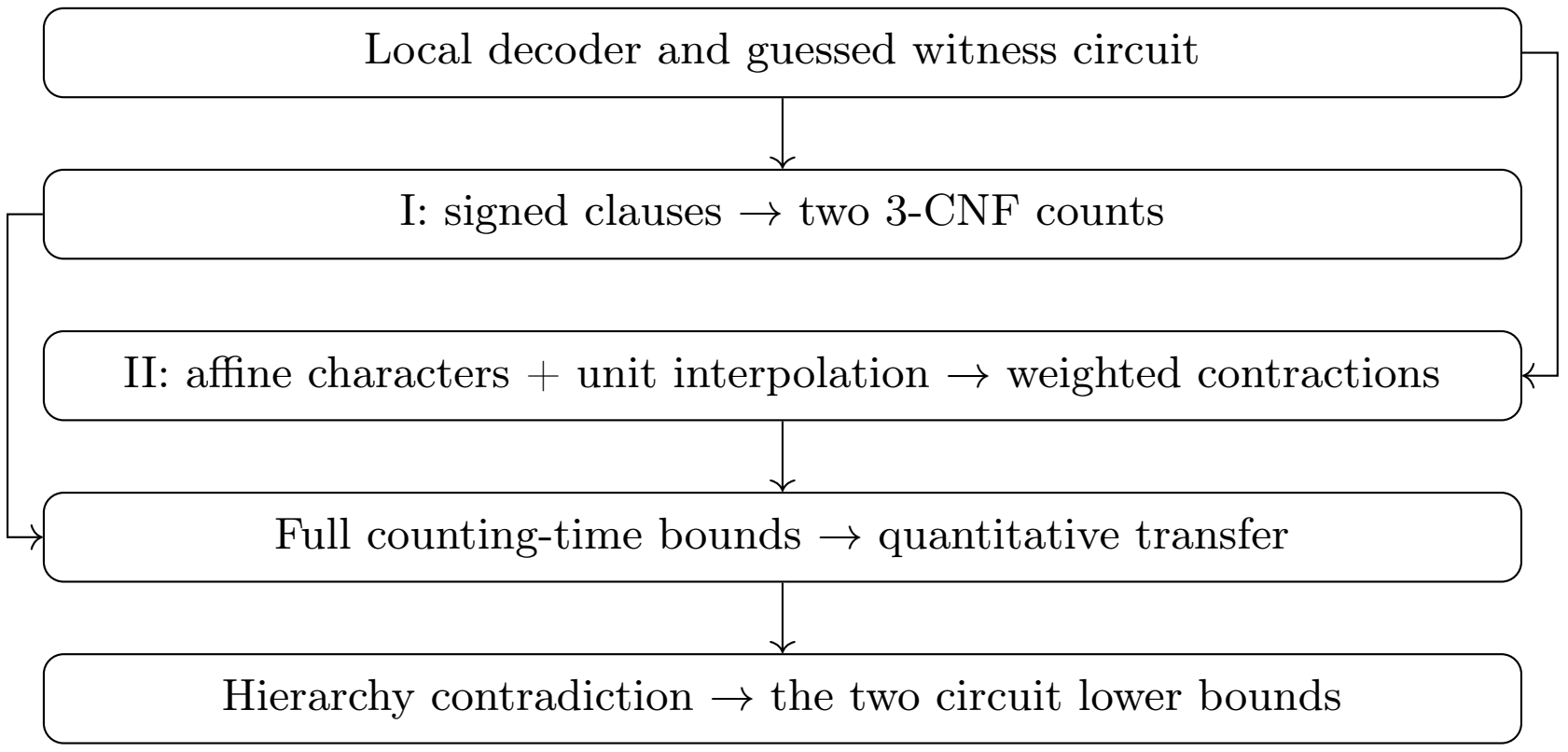

\paragraph{The transfer starts with a hard unary language.}
A local reduction represents a computation on unary input $1^n$ by a 3-CNF whose clauses can be decoded using $O(n)$ binary gates. The bits of its first satisfying assignment define a language in $\mathsf E^{\mathsf{NP}}$. If that language had small circuits at every sufficiently large relevant length, a verifier could guess one such circuit, make three copies, and count the clause addresses it violates. A zero count certifies satisfaction. The general transfer theorem takes the full counting time as its input, including an additive branch budget in the exponent, and chooses constants so that the resulting verifier contradicts the hierarchy.

\paragraph{The first interface preserves the variable budget.}
An arbitrary-fan-in AND or OR gate has one long defining clause and many short implication clauses. Replacing its long clause by a chain of bounded-width clauses would charge variables to its fan-in. A signed selector instead represents the long-clause indicator as a sum of weights: the selector value one has weight $-1$ and forces every literal false. Summing over that selector gives one when the clause is true and zero otherwise. Each long clause thus contributes one selector. Binary XOR gates already have constant-size 3-CNF definitions.

\paragraph{Two ordinary counts recover the signed count.}
A unique-extension XOR chain records the parity of all negative-weight selectors. Its even and odd cases give two ordinary 3-CNFs, and their model-count difference equals the required circuit count. For $n$ inputs and $g$ gates, each formula has at most $n+3g$ variables; its clauses are polynomial in the full circuit description. The variable-dependent term of Anthropic's counting theorem~\cite{anthropic2026} therefore applies at arbitrary clause density. This gives the counter used for unbounded AND/OR with binary XOR, and the transfer yields an $\Omega(m\sqrt{\log m}/\log\log m)$ lower bound.

\paragraph{The second interface pays a logarithmic cost per symmetric gate.}
An ordinary bounded-fan-in encoding of a high-arity symmetric gate computes its input Hamming weight and then evaluates a truth table. That representation charges auxiliary gates to the arity. The intended budget instead charges $\log_2(d_j+1)$ for a gate of arity $d_j$. Unit-separated interpolation achieves this budget by expressing its truth table as a sum of $d_j+1$ products of unary weights. Selecting one summand per gate gives $\prod_j(d_j+1)$ branches. Each branch has the original bounded-arity constraints and accumulated unary weights, so its network size is independent of those symmetric input lists except through polynomial-time weight accumulation.

\paragraph{Affine characters handle arbitrary-fan-in XOR.}
An affine subsystem of rank $r$ has an exact character expansion with $2^r$ terms. Each term contributes unary signs, and the complete sum is $2^r$ times the desired count. Evaluating modulo $2^{N+r+1}$ retains the whole numerator when there are $N$ Boolean variables. The algorithm takes the canonical integer residue and divides by $2^r$. This treats long XOR relations without adding a chain proportional to their input occurrences. The total branch exponent is $r+\sum_j\log_2(d_j+1)$.

\paragraph{Extension-ring arithmetic supports the weighted branches.}
The interpolation coefficients live in a finite unramified extension of $\mathbb Z/2^p\mathbb Z$, where residue-distinct points have unit differences. The contraction theorem used here follows the proof of OpenAI's Theorem D.1~\cite{source} over that coefficient ring. If the input residue field is too small for its interpolation grid, a tower extension provides enough points with logarithmic total degree. Appendix~\ref{app:extension} checks the ring arithmetic, coordinatewise precision, recursion identity, and recovery of the original coordinates. This supplies a deterministic polynomial-space counter with time $2^{B+r+O(s/\sqrt{\log s})}$ up to polynomial factors.

\paragraph{The lower bounds are the final compositions.}
The first counter enters the transfer with the slowly growing denominator $\sqrt{\log s}/\log\log s$. The second enters with $\sqrt{\log s}$ and a branch budget linear in the witness input length. Three witness copies multiply that budget by at most three. The same transfer theorem then gives the two stated lower bounds, with their gate models and budgets carried explicitly. Figure~\ref{fig:overview} displays this common argument and the distinct counting inputs.

\subsection{Related work}
Williams~\cite{williams} develops the algorithm-to-lower-bound connection;
Jahanjou--Miles--Viola~\cite{jmv} give the local reductions used in its
quantitative form. The bounded-binary transfer with the present counting
exponents appears in~\cite[Appendix C.1]{source}. The circuit results here
extend its gate interfaces. Signed large-clause reductions are developed
in~\cite{signedgraph,signedsimple}. Tamaki~\cite{tamaki2016} gives counting
algorithms and $\mathsf E^{\mathsf{NP}}$ lower bounds for depth-two circuits
with symmetric and threshold gates; our symmetric-gate statement allows
arbitrary depth under the stated exceptional-gate budget.
Claude (Anthropic)~\cite[Theorem 1]{anthropic2026} proves the counting
bound used in our first interface. Its clause-dependent term also gives
$2^{O(N/\sqrt{\log N})}$ time for fixed-width CNFs with $O(N)$ clauses.
The corresponding exact-counting statement is formalized in Lean 4
\cite[Appendix E]{anthropic2026}.

\subsection{Organization}
Section~\ref{sec:prelim} states the external counting theorem and the extension-ring contraction theorem. Section~\ref{sec:transfer} proves the common quantitative transfer. Sections~\ref{sec:counter-one} and~\ref{sec:symmetric-gates} construct the two counters. Section~\ref{sec:lower-proofs} composes them with the transfer to prove the main results. Appendix~\ref{app:extension} derives the coefficient-ring extension from the proof of D.1.
\section{Preliminaries and counting theorems}\label{sec:prelim}
All circuits are acyclic, and size counts noninput gates. The full input
description length includes gate input lists and symmetric truth tables.
Logarithms in counting exponents are base two; $\ln$ is the natural
logarithm. We use $\mathsf E^{\mathsf{NP}}=\mathsf{DTIME}(2^{O(m)})^{\mathsf{NP}}$.

\begin{theorem}[{Claude (Anthropic), Theorem 1~\cite{anthropic2026}}]
\label{thm:counting}
There are one deterministic finite-alphabet multi-stack Turing machine
and one absolute constant $C$ such that, for every $k\ge1$, every
$N\ge4$, and every $k$-CNF in which all $N$ variables occur, with
$M\ge2$ distinct clauses, the machine computes its exact model count in
at most
\begin{equation}\label{eq:anthropic-count}
 2^{C\min\{k^2N\log_2\log_2N/\sqrt{\log_2N},\,
                 kM/\sqrt{\log_2M}\}}
\end{equation}
steps, up to integer rounding of the time bound.
\end{theorem}
The cited Theorem 1 appears on pages 1--2 of~\cite{anthropic2026}; Appendix E
specifies the Lean statement and its machine model. A fixed multi-stack
machine has a constant-overhead simulation by a fixed multitape machine.
Repeated clauses can be removed in input-polynomial time. Unused
variables contribute a power of two to the count, and inputs with
$N<4$ or $M<2$ are counted directly.
For $k=3$ the variable term of~\eqref{eq:anthropic-count} gives
$2^{O(N\log\log N/\sqrt{\log N})}$ time at arbitrary clause density.
For fixed $k$ and $M\le DN$, the clause term gives
$2^{O_{k,D}(N/\sqrt{\log N})}$ time. These consequences use the
absolute constant in Theorem~\ref{thm:counting}.

For the second interface, let
\[
 R_p=(\mathbb Z/2^p\mathbb Z)[U]/(f(U)),
\]
where $f$ is a monic binary lift of an irreducible polynomial of degree
$e$ over $\mathbb F_2$.
\begin{theorem}[{Extension-ring form of the contraction theorem of OpenAI}]\label{thm:contraction}
A bipartite Boolean tensor network with $O(s)$ wires, polynomially many
vertices, maximum degree at most $\lceil s^{9/10}\rceil$, and description
length $\mathcal L$ can be contracted over $R_p$ in deterministic time
\[
 (s+\mathcal L)^{O(1)}2^{O(s/\sqrt{\log(s+2)})}
\]
and polynomial space, for $p=O(s)$ and $e=O(\log(s+\mathcal L))$,
when each local entry is computable in
$(s+\mathcal L)^{O(1)}2^{O(s^{9/10})}$ time and polynomial space.
The constants depend on the fixed bounds in these assumptions.
\end{theorem}
The statement follows from the contraction argument in OpenAI~\cite[Appendix D.4]{source}. Appendix~\ref{app:extension} gives the coefficient-ring extension, precision analysis, and recovery of the original coordinates.

\section{A quantitative local-reduction transfer}
\label{sec:transfer}
The transfer is parameterized by the complete test-time bound $T(U,R)=\poly(n)2^{R+AU/a(U)}$. The function $a$ measures the counting speedup and $R$ is the branch exponent. The transfer uses the unary
nondeterministic hierarchy stated in Williams, Theorem 2.1 of the full
version~\cite{williams}, and Jahanjou--Miles--Viola, Theorem 1~\cite{jmv}.
The bounded-binary combination appears in~\cite[Appendix C.1]{source}.

\begin{lemma}[{Uniform linear-size clause decoder}]\label{lem:decoder}
There is a fixed unary language
\[
H_0\in\mathsf{NTIME}(2^n)\setminus\mathsf{NTIME}(2^{n/2})
\]
and a
polynomial-time construction of a decoder with $O(n)$ binary gates
which maps $q=n+O(\log n)$ bits to the three variable addresses and
signs of a 3-CNF $\Phi_n$. The formula is satisfiable exactly when
$1^n\in H_0$. Its variable addresses can be padded to $dn$ bits for
one fixed integer $d$.
\end{lemma}
\begin{proof}
The proof applies the local reduction to a unary hierarchy language and uses the constant support of each decoder output. The hierarchy applies since $2^{(n+1)/2}=o(2^n)$. We simulate the fixed
machine with allowance $T=2^n n^{O(1)}$. We apply the local reduction with
$r=2$, giving $q=\max(\log_2T,n/2)+O(\log n)+O(\log\log T)$.
Each of its $O(n)$ output bits has a constant-depth decision tree and
thus a support of at most a fixed constant $w$ input bits.
The circuit delivered by the reduction can be evaluated in polynomial
time. We evaluate it on all strings of Hamming weight at most $w$.
For any essential coordinate $i$, take a pair of inputs differing only
at $i$ on which the output differs. Zeroing all coordinates outside the
support preserves that difference and puts both inputs among these
queries. Comparing such queried pairs identifies the entire support;
the same queries give its truth table. Realizing each constant-size
truth table takes $O(1)$ binary gates. This constructs the linear-size
decoder uniformly, even if the initial circuit was larger. Bounded
exceptional lengths can be handled separately.
\end{proof}

We declare all $2^{dn}$ variables, including unused addresses. For inputs
$i$ of length $dn$, let $W(i)$ be bit $i$ of the lexicographically first
satisfying assignment of $\Phi_n$, or zero when none exists. We set $W=0$
at all other lengths. The input length determines $n$.
Writing $\Phi_n$ and querying an NP oracle about successive assignment
prefixes takes at most $2^{dn}$ queries of size $2^{O(n)}$. Therefore
$W\in\mathsf E^{\mathsf{NP}}$. 

\begin{theorem}[{Explicit exponent and budget transfer}]\label{thm:transfer}
Let $a$ be eventually nondecreasing, tend to infinity, and satisfy $a(n)=n^{o(1)}$,
$a(cna(n))/a(n)\to1$ for every constant $c>0$, and
$a(dn)/a(n)\to1$ for fixed $d>0$. Suppose a circuit model has polynomial
description length for circuits with $n^{1+o(1)}$ gates, includes binary
gates, and admits the following deterministic test for a guessed witness
circuit $C$ of $s$ gates on $dn$ bits. Form three copies with the decoder
of Lemma~\ref{lem:decoder} and count violating clause addresses. Suppose
this test has a deterministic multitape implementation taking time at most
\[
 \poly(n)2^{R+A U/a(U)},\qquad
 U=\kappa s+Bn,\qquad R\le b\delta dn,
\]
for fixed positive $A,B,\kappa,b$, whenever the permitted witness
budget is at most $\delta dn$. The exponent bound must include all
preprocessing, branches and exact reconstruction. Assume an effective
integer threshold $s_n=(\eta+o(1))na(n)$ can be computed in polynomial time.
Then for sufficiently small positive $\delta,c$, at infinitely many
lengths $m$ every permitted circuit computing $W$ has more than
$cma(m)$ gates. For a model without a budget take $R=0$ and omit $\delta$.
\end{theorem}
\begin{proof}
The proof guesses a small witness circuit and uses the assumed counting time to contradict the unary nondeterministic hierarchy. Choose $\delta$ so $b\delta d<1/8$ and $\eta$ so $A\kappa\eta<1/8$.
If the conclusion failed, sufficiently small $c<\eta/d$ and the
fixed-scale regularity of $a$ would provide witness circuits of at most
$s_n$ gates at every sufficiently large address length $dn$.
On input $1^n$, guess such a circuit and form the violation circuit.
All guessing, budget checks and decoding use polynomial time.
We accept precisely when the number of violated clause addresses is zero.
For a satisfiable $\Phi_n$, its first-witness circuit is an accepting
guess. For an unsatisfiable $\Phi_n$, every guessed circuit describes an
assignment violating at least one clause. Soundness holds for every guessed circuit.

Now $U=(\kappa\eta+o(1))na(n)$. We squeeze between fixed positive
multiples of $na(n)$; monotonicity and regularity give
$AU/a(U)=(A\kappa\eta+o(1))n$.
The total verifier time is $2^{(b\delta d+A\kappa\eta+o(1))n}$,
less than $2^{n/2}$ for all sufficiently large $n$.
The stipulated multitape bound includes all workspace access costs.
We hardwire the finitely many smaller lengths and reject nonunary inputs.
This contradicts the chosen hierarchy language. Thus the bound fails
at infinitely many address lengths $dn$, which proves the result.
\end{proof}

For sufficiently large $x$, the two applications use
\[
 a_0(x)=\sqrt{\log_2x},\qquad a_1(x)=\frac{\sqrt{\log_2x}}{\log_2\log_2x}.
\]
Their regularity follows from $\log(cna_i(n))=\log n+O(\log\log n)$.
For $a_0$ take\[s_n=\lfloor\eta n\lfloor\sqrt{\lfloor\log_2n\rfloor}\rfloor\rfloor.\]
For $a_1$, divide the inner expression by
$\lceil\log_2\lfloor\log_2n\rfloor\rceil$ before the outer floor.
These thresholds use integer arithmetic and have the required asymptotics.

\begin{center}\small
\setlength{\tabcolsep}{3pt}\begin{tabular}{p{.30\linewidth}p{.27\linewidth}p{.34\linewidth}}
\toprule
Witness model & Counting cost on the violation instance & Consequence \\
\midrule
Unbounded AND/OR; binary XOR & $R=0$, $U=9s+O(n)$, $a=a_1$ &
$\Omega(m\sqrt{\log m}/{}$ $\log\log m)$ gates \\
Binary base; budgeted arbitrary XOR and other symmetric gates &
$R\le3\delta dn$, $U=3s+O(n)$, $a=a_0$ &
$\Omega(m\sqrt{\log m})$ gates under the budget \\
\bottomrule
\end{tabular}
\end{center}
\section{Counting algorithm I: signed long clauses}\label{sec:counter-one}
This counter preserves a linear variable budget by replacing each long defining clause with one signed selector. An auxiliary parity chain converts the resulting signed sum into two ordinary 3-CNF counts. Lemma~\ref{lem:unbounded-count} supplies the complete bound needed by the transfer.

The circuit class $\mathcal U$ is defined in Section~1.2. 

\begin{lemma}\label{lem:unbounded-count}
For a circuit in $\mathcal U$ with $n$ inputs, $g$ gates and full
description length $L$, its output-one preimages are the difference
of the model counts of two ordinary 3-CNFs, each on at most
$V=n+g+2H\le n+3g$ variables and $O(L+g)$ clauses. Here $H$ is the
number of AND/OR gates of fan-in at least three. By Theorem~\ref{thm:counting}, the two counts are computable in time
\[
 (L+V)^{O(1)}2^{O(V\log\log(V+4)/\sqrt{\log(V+2)})}.
\]
\end{lemma}
\begin{proof}
The proof replaces long defining clauses by signed selectors and separates their total sign with one parity chain. The construction uses one variable per gate. An AND definition uses all
$(\neg z\vee l_i)$ and the single clause
$(z\vee\bigvee_i\neg l_i)$; an OR definition uses
$(z\vee\neg l_i)$ and $(\neg z\vee\bigvee_i l_i)$.
XOR has four ternary clauses and COPY has two binary clauses.
There is a unique gate extension of each input. For a long clause
$\bigvee_i l_i$, replace it with $(\neg s\vee\neg l_i)$ and give
$s=1$ weight $-1$. With all input and gate variables fixed, the
selector sum is one for a true clause and zero for a false clause.
This is the existing signed-clause construction
\cite{signedgraph,signedsimple}. A unique-extension XOR chain on
the $H$ selectors uses $H-1$ further variables. We fix its parity to
zero and one to obtain the two formulas; for $H=0$, use an
inconsistent second formula. The count difference is exactly the
claimed number. Each individual count needs at most $V+1$ bits.
After duplicate clauses are removed, let $V_i$ be the number of occurring
variables of the $i$th formula, and let $M_i$ be its clause count.
Theorem~\ref{thm:counting}, with $k=3$, gives
\[
 2^{C\min\{9V_i\log_2\log_2V_i/\sqrt{\log_2V_i},\,
                    3M_i/\sqrt{\log_2M_i}\}}.
\]
Restoring the $V-V_i$ unused variables multiplies its output by
$2^{V-V_i}$. The first term, $V_i\le V$, and polynomial-time
normalization give the displayed general-density bound. The cases
$V_i<4$ or $M_i<2$ are evaluated directly.
\end{proof}

\section{Counting algorithm II: affine characters and unit interpolation}\label{sec:symmetric-gates}
The second counter removes affine relations by characters and arbitrary symmetric gates by a separated interpolation identity. Both transformations produce unary weights. Theorem~\ref{thm:symmetric-counter} combines their branch count with the extension-ring contraction theorem and reconstructs the exact integer count.
\subsection{Exact affine-character removal}
\label{sec:affinecharacters}
The following identity supplies the XOR part of the symmetric-gate
counter. On $N$ Boolean variables, row-reduce $Ax=a$ over $\F$.
An inconsistent system contributes zero. Otherwise retain $r$ independent
rows $Rx=d$. For every $x$,
\[
 2^r[Rx=d]=\sum_{y\in\F^r}(-1)^{y\cdot d}\,
                                (-1)^{(R^Ty)\cdot x}.
\]
This follows by factoring the character sum into $r$ sums over single
bits; a nonzero residual makes one such factor zero. Multiplication by
any other constraint product and summation over $x$ proves the identity
for its count. Every character becomes a unary sign on each variable.
In a coefficient ring of characteristic $2^p$, perform all branch sums
at precision $p=N+r+1$. If the final unscaled count is an integer $Z$
with $0\le Z\le2^N$, the resulting canonical integer residue is exactly
$2^rZ<2^{N+r+1}$. We divide this integer by $2^r$ afterwards.
 Polynomially many bits suffice,
and sequential character enumeration uses polynomial space.

For the mixed decomposition below, individual branches can have
extension-ring values. The identity applies before any decomposition,
so the complete sum is the integer $2^rZ$. This identity justifies the final reconstruction.
\subsection{Unit interpolation for symmetric gates}

We apply the unit-separated interpolation construction of
OpenAI~\cite[Appendix D.4.1]{source} before contraction.
It removes arbitrary symmetric gates from an otherwise bounded-arity
circuit and charges their arities through a multiplicative branch count.

\subsubsection{Coefficient ring and interpolation points}
For the explicit input length $\mathcal L$, choose
\[
 e=\max\{4\lceil\log_2s\rceil,
              \lceil\log_2(d_{\max}+1)\rceil,1\},
\]
where $d_{\max}\le\mathcal L$ is the largest symmetric arity.
The irreducible polynomial defining $R_p$ can be found in time polynomial
in $s+\mathcal L$, since $2^e=O(s^4+d_{\max}+1)$.
Ring operations and unit inversions take polynomial time in $e,p$.
Binary-coordinate representatives give $2^e$ points with pairwise unit
differences. Theorem~\ref{thm:contraction} supplies contraction for
the resulting $R_p$-valued branches, with this coefficient ring fixed
throughout each call.

\subsubsection{An integral separated decomposition}
Let a gate have $d$ input occurrences, with Boolean truth table
$f(0),\ldots,f(d)$ indexed by input Hamming weight. Its output bit
is $y$. We choose distinct residue-field lifts
$\alpha_0,\ldots,\alpha_d$ in $R_p$; zero is allowed.
The matrix $V_{t,j}=\alpha_j^t$, $0\le t,j\le d$, has determinant
$\prod_{i<j}(\alpha_j-\alpha_i)$, a unit. For each $y\in\{0,1\}$
solve
\begin{equation}\label{eq:symmetric-vandermonde}
 \sum_{j=0}^d c_{j,y}\alpha_j^t=[y=f(t)],\qquad 0\le t\le d.
\end{equation}
These two solves take polynomial time in $d,e,p$ and lose no precision.
Elimination always has a unit pivot after a row exchange: reduction
modulo two is an invertible matrix over a field. Equivalently the
coefficients are defined integrally in $\mathcal O$ and reduced
modulo $2^p$.

For any Boolean input occurrences $b_1,\ldots,b_d$, the exact identity is
\begin{equation}\label{eq:symmetric-separation}
 [y=f(\textstyle\sum_i b_i)]
     =\sum_{j=0}^d c_{j,y}\prod_{i=1}^d\alpha_j^{b_i}
       \quad\hbox{in }R_p.
\end{equation}
We use $0^0=1$. A positive literal on variable $x$ contributes unary
weights $(1,\alpha_j)$, and a negative literal contributes
$(\alpha_j,1)$. The output variable receives weights
$(c_{j,0},c_{j,1})$. If a variable occurs more than once, multiply
all its unary weights. Shared inputs, fanout and negated inputs
therefore cause no independence assumption. When the gates have
arbitrary depth, an output can receive both its own coefficient
weight and factors from later gates; multiplication still applies
pointwise. 

\subsection{A complete weighted counting interface}
We consider $N$ Boolean variables, $m_0$ hard predicates of fixed arity,
arbitrary explicitly described symmetric gate relations of arities
$d_1,\ldots,d_h$, and an affine subsystem of rank $r$. Let
\[
 s=N+m_0+2,\qquad B=\sum_{j=1}^h\log_2(d_j+1).
\]
Inconsistent affine systems give count zero. Otherwise the exact
character identity from Section~\ref{sec:affinecharacters} removes
the $r$ independent equations using $2^r$ terms and an eventual
integer division by $2^r$.

\begin{theorem}\label{thm:symmetric-counter}
The preceding mixed system
has a deterministic exact counting algorithm with time
\begin{equation}\label{eq:symmetric-time}
 (s+\mathcal L)^{O(1)}\,
       2^{B+r+O(s/\sqrt{\log(s+2)})}
\end{equation}
and polynomial space. Constants depend only on the hard-predicate
arity. Recovery and original-assignment unranking use a polynomial number
of deterministic counting calls. When the count is positive, exact uniform
sampling uses a polynomial number of counting calls and polynomial expected
additional time with fair random bits.
\end{theorem}
\begin{proof}
The proof expands the affine and symmetric constraints into unary weights, contracts each branch, and reconstructs one exact integer. The precision is $p=N+r+1\le2N+1$. We enumerate one interpolation index for each
symmetric gate and one affine character, sequentially. There are
exactly $2^r\prod_j(d_j+1)=2^{B+r}$ branches. A branch has only
the fixed-arity hard predicates and a pair of unary weights
$w_v(0),w_v(1)\in R_p$ at each original variable. All these
weights can be accumulated in input-polynomial time and space.

We make the fixed-density network of source D.2: one tensor per hard
predicate, equality trees for its variable occurrences, and the
unary weights attached at one port of each variable. A variable
with no hard occurrences contributes $w_v(0)+w_v(1)$ as a scalar.
There are $O(s)$ wires, bounded maximum degree, and polynomial-time
local entries. Subdivision makes the graph bipartite without changing
its contraction. Theorem~\ref{thm:contraction} evaluates the
branch in the structural time in \eqref{eq:symmetric-time}, with
polynomial space. Each symmetric input occurrence contributes to an accumulated unary weight.

We sum all branch residues with their affine-character signs.
By the two pointwise identities the result is the constant-coordinate
integer $2^rZ$ modulo $2^{N+r+1}$, where $0\le Z\le2^N$ is the
required count. All other coordinates vanish. The canonical constant
residue is the actual numerator, so divide it by $2^r$ as an integer.
In the absence of affine equations $p=N+1$ suffices, regardless of
the symmetric arities. 

We pin original variables successively with unary hard predicates.
The resulting instances have $N$ variables, the same symmetric decompositions and affine rank bound, and at most $N$ additional hard predicates. The child counts sum
to their parent's count. The usual count self-reduction recovers a
solution and maps each integer in $[0,Z)$ bijectively to one solution.
For exact uniform sampling when $Z>1$, let $b=\lceil\log_2 Z\rceil$.
A trial draws $b$ independent fair bits, interprets them as an integer
$a\in[0,2^b)$, and accepts precisely when $a<Z$. The accepted index is
uniform on $[0,Z)$, and the acceptance probability $Z/2^b>1/2$ gives
fewer than two trials in expectation. For $Z=1$, the index is zero.
Unranking the accepted index therefore yields an exact uniform solution
with $O(\log(Z+1))$ expected additional time. We stream all branch choices
and keep one accumulator, proving the space bound.
\end{proof}

For an acyclic circuit with $n$ inputs and $g$ gates, give every gate
one output variable, so $N=n+g$. Every bounded binary gate imposes
one arity-three predicate. We remove selected arbitrary-fan-in XOR gates
by characters and all other designated symmetric gates by
\eqref{eq:symmetric-separation}; pin the requested output.
Then $m_0=O(N)$ and \eqref{eq:symmetric-time} becomes
\[
 \poly(\mathcal L+N)2^{B+r+O(N/\sqrt{\log(N+2)})}.
\]
Each input has a unique full gate assignment, so the answer counts
original input preimages.
For example, $B+r=O(N/\sqrt{\log N})$ preserves the fixed-density
exponent, while $B+r=o(N)$ gives subexponential time in $N$.
These statements have no depth or fanout restriction.
They include majority and exact-Hamming-weight gates.

\section{Proofs of the circuit lower bounds}\label{sec:lower-proofs}
Both conclusions follow from Theorem~\ref{thm:transfer}. The first proof substitutes the two-CNF counter with denominator $a_1$; the second substitutes the weighted counter with denominator $a_0$ and the explicit exceptional-gate budget. The witness language is the one constructed in Section~\ref{sec:transfer}.
\subsection{Unbounded AND/OR and binary XOR}
\begin{proof}[Proof of Theorem~\ref{thm:unbounded-lower}]
We apply Theorem~\ref{thm:transfer} with $a=a_1$ and no budget.
A witness circuit with $s$ gates on $dn$ bits has three copies in
the violation circuit, along with the $O(n)$-gate binary decoder.
There are $q=n+O(\log n)$ inputs, at most $3s+B_0n$ gates and at
most $3s+1$ long defining clauses. Lemma~\ref{lem:unbounded-count}
therefore yields two 3-CNFs on at most $9s+Bn$ variables.
Each has polynomially many clauses: duplicate equal literals can be
removed from AND/OR gates, so a complete witness description has length
$O(s(dn+s)\log(dn+s+2))$. The three-copy decoder, output pin and
parity definitions add only polynomial description length.

The complete verifier time is
\[\poly(n)2^{A U/a_1(U)},\qquad U=9s+Bn.\] One may pad both
formulas with the same number of unused variables; this multiplies
both counts by the same power of two and preserves zero difference.
The exact difference is the nonnegative number of violated clause
addresses, so the transfer's completeness and soundness apply.
For example, any sufficiently small rational $\eta<1/(72A)$ meets
its time requirement. The two calls run sequentially; the final subtraction uses $O(U)$ bits.
\end{proof}
\subsection{A quantitative larger-basis circuit lower bound}
We use the standard convention that a gate input list contains no
duplicate signed wires, so $d_j\le2(m+g)$ on $m$ inputs and $g$
gates. Explicitly repeated lists use their actual $B$ in the counting algorithm.
We give arbitrary-fan-in XOR gates unit budget cost and all other
designated symmetric gates cost $\log_2(d_j+1)$.
Let $b(C)$ be the sum of these costs; ordinary binary gates are free
in this budget but still count toward $g$.

\begin{proof}[Proof of Theorem~\ref{thm:symmetric-lower}]
We use Theorem~\ref{thm:transfer} with $a=a_0$. A witness circuit on
$dn$ bits with $s$ gates and $b(C)\le\delta dn$ produces a violation
circuit with $N\le3s+B_0n$ input and gate variables. The decoder,
sign checks and final AND use bounded binary gates. The three copies
have combined removed cost $B+r\le3\delta dn$: the affine rank is
at most the number of removed XOR rows. Thus
Theorem~\ref{thm:symmetric-counter}, with fixed-density hard gates,
gives $\poly(n)2^{3\delta dn+A U/a_0(U)}$ for $U=3s+B_1n$.
The coefficient of the budget term is exactly one; it is the binary
logarithm of the explicit branch count.

Every witness gate has a polynomial-size input list and, for a
symmetric gate, a table of $d_j+1$ bits. After substitution of decoded
addresses, occurrences that coincide are retained. The counter already
supports repeated inputs with the original witness budget.
Polynomial space gives a multitape simulation with polynomial overhead.
The transfer theorem applies with $3\delta d<1/8$ and $3A\eta<1/8$.
The constants are independent of the gate arities.

For its final consequence it suffices to consider $g<c m\sqrt{\log m}$,
since larger circuits already meet the conclusion. For all sufficiently
large $m$, $\log_2(2(m+g)+1)\le2\log_2(m+2)$.
The stated gate allowances therefore have budget at most
$\epsilon m+2\epsilon m=3\epsilon m$. We take $3\epsilon\le\delta$.
\end{proof}
\section{Discussion and open problems}
The two lower bounds extend the binary-gate conclusion of
OpenAI~\cite[Corollary 1.4 and Appendix C.1]{source} through distinct
counting interfaces. Signed selectors preserve a linear variable budget
for arbitrary-fan-in AND/OR gates. Affine characters and unit-separated
interpolation preserve a linear wire budget after the symmetric gates
have been expanded, with the expansion charged by $B+r$.

An ordinary bounded-fan-in encoding of a symmetric gate of arity $d$
computes its Hamming weight with $O(d\log(d+2))$ auxiliary gates and
evaluates its explicit truth table with $O(d)$ further gates. Across all
exceptional gates this gives a CNF size controlled by
$N+\sum_j d_j\log(d_j+2)$, whereas the theorem budgets
$\sum_j\log_2(d_j+1)$. A CNF counting reduction with the latter budget
would give a counting-theorem basis for the symmetric-gate bound.
The present proof uses Theorem~\ref{thm:contraction} for its
extension-ring branches.

Two further questions are whether the logarithmic loss in the unbounded
AND/OR lower bound can be removed, and whether a larger exceptional-gate
budget permits the square-root-logarithmic lower bound.

\section*{AI assistance}
OpenAI Codex was used to assist with drafting and revising the manuscript,
developing parts of the mathematical arguments, and checking derivations.

\bibliographystyle{splncs04}
\bibliography{refs}
\appendix
\section{Extension-ring contraction from the proof of Theorem D.1}\label{app:extension}
\begin{proof}[Proof of Theorem~\ref{thm:contraction}]
The idea is to run the contraction recursion of OpenAI~\cite[Appendix D.4]{source} over an unramified coefficient algebra containing the input ring. The Boolean network and all recursion parameters are kept fixed. The following construction gives the coefficient algebra and checks the places where that proof uses its ring.

\paragraph{A coefficient algebra containing the input.}
Let $\mathcal O_0=\mathbb Z_{(2)}[U]/(f)$, where $f$ is the monic lift in the theorem. It is free of rank $e$ over $\mathbb Z_{(2)}$, with residue field $\mathbb F_{2^e}$. Put
\[
 t=\max\{1,\lceil 4\lceil\log_2s\rceil/e\rceil\},\qquad E=et.
\]
Then $2^E\ge s^4$ and $E<e+4\lceil\log_2s\rceil$, except for the harmless equality case $t=1$. If $t=1$, take $\mathcal O=\mathcal O_0$. Otherwise choose a monic degree-$t$ polynomial $\bar g$ irreducible over $\mathbb F_{2^e}$, lift each coefficient to its binary-coordinate representative in $\mathcal O_0$, and write the resulting polynomial as $g$. Set
\[
 \mathcal O=\mathcal O_0[Z]/(g),\qquad A_q=\mathcal O/(2^q).
\]
The monic quotient makes $\mathcal O$ free of rank $t$ over $\mathcal O_0$. Its residue field is $\mathbb F_{2^E}$, and the input ring $R_p=\mathcal O_0/(2^p)$ embeds as the constant-$Z$ coordinates of $A_p$. In particular, this embedding is injective at every precision.

This algebra is constructible within the polynomial factor of the theorem. Enumerating monic degree-$t$ polynomials and testing irreducibility over the explicitly represented finite field takes $2^{O(et)}\poly(e,t)$ bit operations, even using trial division by monic polynomials of degree at most $t/2$. Since $E=O(\log(s+\mathcal L))$, that bound is polynomial in $s+\mathcal L$. The construction uses the same elementary finite-field fact as Appendix D.4.1 of the source.

\paragraph{Units, valuation, and precision.}
Use the free basis $U^iZ^j$, $0\le i<e$, $0\le j<t$, with the evident basis when $t=1$. For a nonzero element, its two-adic valuation is the minimum coordinate valuation. After factoring out this power of two, its reduction in the residue field is nonzero. Products of two such reductions are nonzero, proving multiplicativity of the valuation. An element nonzero modulo two is a unit: its multiplication matrix has odd determinant, whose inverse exists in $\mathbb Z_{(2)}$. These arguments also show that $\mathcal O$ is a domain and that its rational scalar extension is a field.

Arithmetic modulo $2^q$, division by an explicitly known power of two with the corresponding precision loss, and unit inversion all use polynomial time and space in $E,q$ and the coefficient descriptions. In particular, divisibility by $2^a$ is coordinatewise. The $2^E$ binary-coordinate residue lifts have pairwise unit differences and provide every interpolation point required in Appendix D.4.1. The recursion needs fewer than $s^4$ such points.

\paragraph{Entrywise lifts and the deformation.}
At creation precision $q_c$, each entry is represented by the integers in $[0,2^{q_c})$ in the fixed free basis. At every later precision those same coordinates are used. This is precisely the creation-precision convention in Appendix D.4.1. It applies to input entries as well as recursively created packet tensors. Each call therefore has fixed integral lifts of its tensors, and its deformed contraction $F(T)$ is a polynomial over the rational scalar extension of $\mathcal O$.

The packing construction in Appendix D.4.2 depends on the network, not on its coefficients. The Cauchy kernel in Appendix D.4.3 has rational scalar coefficients, and Lemma D.4 proves their integrality after the substitution $T=t_0+\alpha z$ by cancellation of explicit integer powers of two. Consequently the same calculation gives $F(t_0+\alpha z)\in\mathcal O[z]$ for our lifted entries. Here $t_0$ denotes a source interpolation node, distinct from the tower degree $t$ above. The source proof already establishes this step for arbitrary integral coefficients of its extension algebra.

\paragraph{Hermite recovery and mode truncation.}
Lemma D.5 in Appendix D.4.4 uses rational scalar Hermite multipliers. Their two-adic valuations, the jet multiplicities, and the precisions $p_{t_0}$ depend on integer interpolation nodes and the network parameters. Applying that identity in each free coordinate gives exactly the same precision loss and exact recovery modulo the parent's precision. Interpolation from the residue-distinct values is valid because every Vandermonde difference is a unit. All intermediate precision bounds remain those of source equation (D.11).

The mode expansion in Appendix D.4.5 has coefficients divisible by the stated powers of two. Multiplication by arbitrary integral tensor entries preserves that divisibility. Thus precisely the same weighted-degree cutoff discards terms that vanish modulo $2^{p_{t_0}}$. The factorization into child contractions in equation (D.16) is an identity of sums and products in a commutative ring. It applies in $A_{p_{t_0}}$, with no change to the branch index set, child edge counts, or degree bounds.

\paragraph{Local descriptions and total complexity.}
Input entry routines are evaluated at their declared creation precision and embedded into the constant-$Z$ coordinates. This adds a polynomial factor in $E$ to their stated cost. New packet routines are the routines of Appendix D.4.6, evaluated in the fixed basis of $\mathcal O$; their cost remains $\poly(s+\mathcal L)2^{O(s^{9/10})}$ with polynomial space. The descriptions and their creation precisions are shared as in the source recursion.

The node-count estimate of Appendix D.4.7 is independent of the coefficient degree. It gives $2^{O(s/\sqrt{\log s})}$ calls, and the direct leaves and local-entry costs fit the same exponent. Arithmetic contributes a polynomial in $E$, the precisions, and $s+\mathcal L$. Since $E=O(\log(s+\mathcal L))$ and all precisions are polynomial in $s$, the total bound is the one stated in Theorem~\ref{thm:contraction}. Depth-first traversal and coordinatewise accumulators give polynomial working space.

\paragraph{Returning to the input ring.}
The root contraction is a polynomial with integer coefficients in its original entries. Its value therefore lies in the embedded $R_p$. Reading the constant-$Z$ coordinates returns it; the other tower coordinates are zero modulo $2^p$. No additional interpolation, division, or loss of precision is needed for this final extraction. This proves the extension-ring statement from the contraction argument of Theorem D.1.
\end{proof}

\end{document}